\documentclass[fleqn,usenatbib]{mnras}
\usepackage{newtxtext,newtxmath}
\usepackage[british]{babel}
\usepackage{multicol}
\usepackage{dblfloatfix}

\usepackage[T1]{fontenc}
\usepackage{ae,aecompl}

\usepackage{amsmath,amstext}
\usepackage[all]{hypcap}
\usepackage{newtxmath}
\usepackage{afterpage}
\usepackage{gensymb}
\usepackage{enumitem}
\usepackage[colorinlistoftodos]{todonotes}
\usepackage{url}
\usepackage{graphicx}
\usepackage{amssymb}
\usepackage{color}

\hypersetup{linkcolor=cyan,citecolor=violet,filecolor=cyan,urlcolor=blue}

\usepackage{soul}

\title[MWC 560]{Broad absorption line symbiotic stars: \\highly ionized species in the fast outflow from MWC 560}

\author[Lucy et al.]{
Adrian B. Lucy,$^{1}$\thanks{lucy@astro.columbia.edu}\thanks{LSSTC Data Science Fellow}
Christian Knigge,$^{2}$
and J. L. Sokoloski$^{1}$
\\
$^{1}$Columbia University, Dept. of Astronomy, 550 West 120th Street, New York, NY 10027, U.S.A.\\
$^{2}$University of Southampton, School of Physics \& Astronomy, Highfield, Southampton, SO17 1BJ, U.K.
}

\date{}

\pubyear{2018}

\begin{document}
\label{firstpage}
\pagerange{\pageref{firstpage}--\pageref{lastpage}}
\maketitle

\begin{abstract}
In symbiotic binaries, jets and disk winds may be integral to the physics of accretion onto white dwarfs from cool giants. The persistent outflow from symbiotic star MWC~560 ($\equiv$V694~Mon) is known to manifest as broad absorption lines (BALs), most prominently at the Balmer transitions. We report the detection of high-ionization BALs from \ion{C}{iv}, \ion{Si}{iv}, \ion{N}{v}, and \ion{He}{ii} in {\it International Ultraviolet Explorer} spectra obtained on 1990~April~29\,$-$\,30, when an optical outburst temporarily erased the obscuring `iron curtain' of absorption troughs from \ion{Fe}{ii} and similar ions. The \ion{C}{iv} and \ion{Si}{iv} BALs reached maximum radial velocities at least 1000~km~s$^{-1}$ higher than contemporaneous \ion{Mg}{ii} and \ion{He}{ii} BALs; the same behaviors occur in the winds of quasars and cataclysmic variables. An iron curtain lifts to unveil high-ionization BALs during the P~Cygni phase observed in some novae, suggesting by analogy a temporary switch in MWC 560 from persistent outflow to discrete mass ejection. At least three more symbiotic stars exhibit broad absorption with blue edges faster than 1500~km~s$^{-1}$; high-ionization BALs have been reported in AS~304 ($\equiv$V4018~Sgr), while transient Balmer BALs have been reported in Z~And and CH~Cyg. These BAL-producing fast outflows can have wider opening angles than has been previously supposed. BAL symbiotics are short-timescale laboratories for their giga-scale analogs, broad absorption line quasars (BALQSOs), which display a similarly wide range of ionization states in their winds.
\end{abstract}





\section{Introduction} \label{intro}

\begin{figure*}
\includegraphics[width=7in]{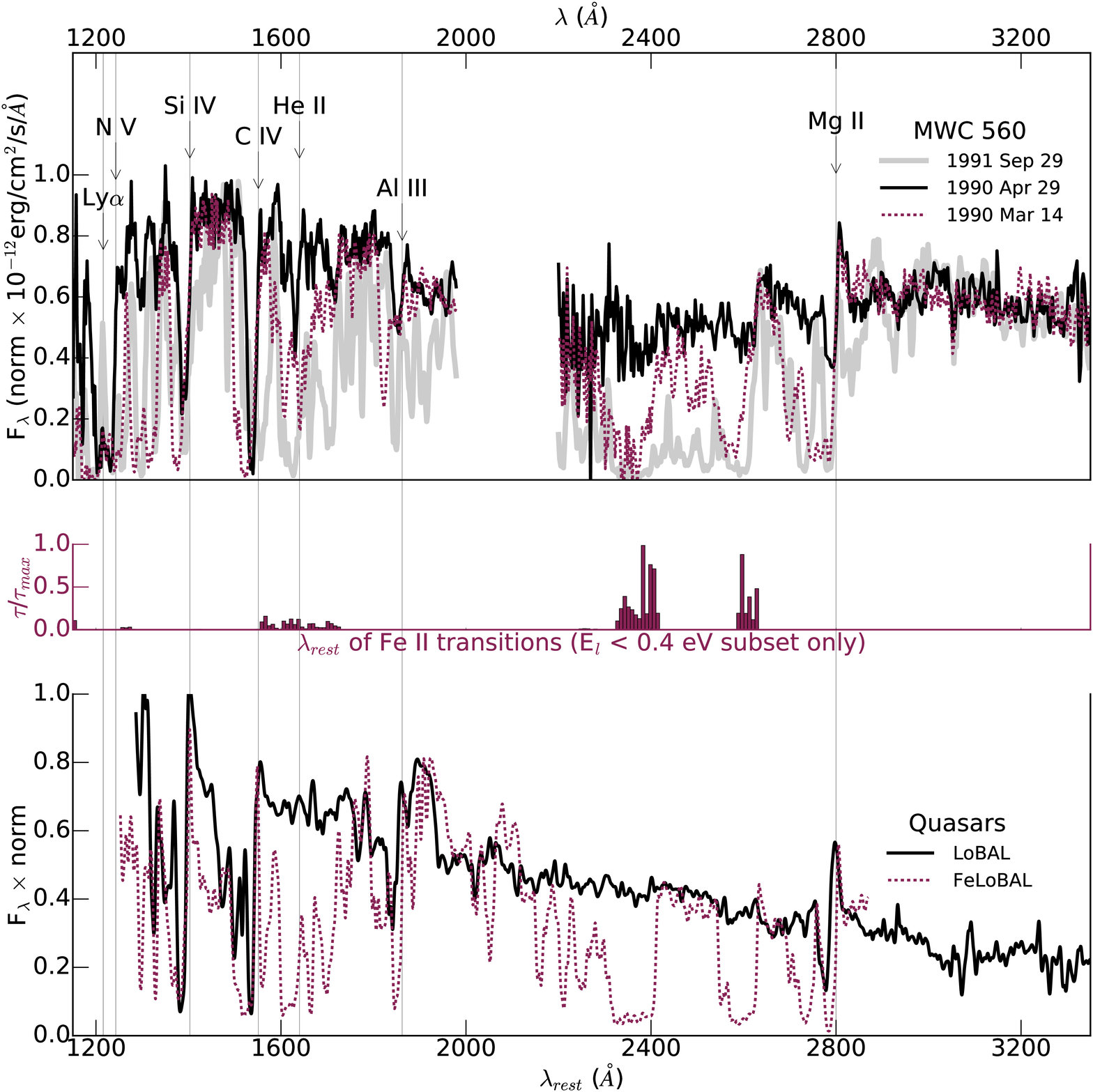}
\caption{\protect\\ {\it Top panel:} IUE spectra show that MWC 560's iron curtain obscured the UV in 1990 March, vanished in 1990 April, and returned by 1991 September. We re-normalized all spectra to the maximum (1990 April 29) median flux density longward of 3000\AA, and did not deredden. Line identifications for the 1990 April spectrum are marked at their rest wavelengths. \protect\\ {\it Middle panel:} Predicted optical depth ($\tau$) relative to the strongest value ($\tau_{max}$), for the \ion{Fe}{ii} transitions from near the ground state (\autoref{3.2}) in 10\AA\ bins, excluding higher-excitations populated during iron curtain phases (e.g., on the blue wing of \ion{Mg}{ii}). \protect\\ {\it Bottom panel:} The Balmer FeLoBAL quasar J172341.1+555340.5 \citep{Hall2002} resembles the iron curtain states of MWC 560, while the LoBAL quasar J140125.6+581650.7 \citep{gibson2009} resembles the 1990 April spectrum (\autoref{4.3}). These Sloan Digital Sky Survey spectra were obtained from the Science Archive Server, corrected for redshift and the negligible Galactic extinction on their sightlines \citep{green2015}, normalized to match pseudo-continua on an arbitrary scale, and smoothed to resolving power R=400. The FeLoBAL was further dereddened in its rest frame for E(B-V)=0.15 of SMC bar extinction \citep{gordon2003}.} \label{fig1}
\end{figure*}

Accretion onto compact objects exhibits similar phenomenology across a wide range of mass scales, spanning over nine orders of magnitude from supermassive black holes to white dwarfs. In fact, MWC 560 ($\equiv$V694~Mon) and CH Cyg---each believed to constitute a white dwarf (WD) accreting from a red giant (RG), the most common kind of symbiotic star---have been called `nano-quasars,' because broadening their emission lines by a factor of 10 produces optical spectra that almost identically resemble the archetypal Seyfert 1 galaxy I~Zw~1 \citep{Zamanov2002}. 

MWC 560 also usually exhibits blue-shifted broad absorption lines (BALs\footnote{We ignore the distinction between BALs and mini-BALs appearing in the quasar context, where the richer sample size sometimes motivates formalism \citep[e.g.,][]{knigge2008}.}) from a jet \citep{Schmid2001} or disk wind. The blue edges of these lines vary between 900--6000\,km\,s$^{-1}$ from the Balmer transitions and combinations of \ion{Al}{iii}, \ion{Mg}{ii}, \ion{Fe}{ii}, \ion{Cr}{ii}, \ion{Si}{ii}, \ion{C}{ii}, \ion{Ca}{ii}, \ion{Mg}{i}, \ion{Na}{i}, \ion{O}{i}, and \ion{He}{i}, including various excited-state, metastable-state, and resonance transitions \citep{Tomov1990,Tomov1992,Michalitsianos1991,Schmid2001,Meier1996}. MWC 560 thus resembles the rare subset of \ion{Fe}{ii} low-ionization BAL quasars \mbox{(FeLoBALs)} with Balmer BALs; symbiotic nebulae typically have electron densities up to 10$^{8}$--10$^{12}$\,cm$^{-3}$, comparable to the densities necessary to populate Balmer absorption in photo-ionized quasar winds \citep{Hall2007,leighly2011,williams2017}. However, MWC 560 has {\it seemed} to be missing the set of higher-ionization ultraviolet (UV) resonance BALs---\ion{C}{iv}\,$\lambda$1550\AA, \ion{Si}{iv}\,$\lambda$1400\AA, and \ion{N}{v}\,$\lambda$1240\AA---exhibited by many cataclysmic variable (CV) winds \citep{Shlosman1993} and all BAL quasars outflows \citep{weymann1991}. 

Optical flickering from the accretion disk, TiO bands from the M giant, and BALs from the disk outflow led \citet{Bond1984} to an accurate description of MWC 560, but the system became popular during an optical brightening peak in 1990. The Balmer BALs reacted dramatically, with blue edges up to 6000~km~s$^{-1}$ shifting by thousands of km~s$^{-1}$ in mere days \citep{Tomov1990}, and the inner accretion disk may have been partially evacuated during that year \citep{Zamanov2011flickstop}. MWC~560 later attracted renewed interest in 2016, when it brightened to V$\approx$9 for the first time since 1990 \citep{Munari2016}. No signs of inner-disk evacuation were observed in 2016---and once the peak optical flux was reached, the Balmer BALs varied stably alongside weeks-long changes in the accretion rate, with blue edges up to 3000 km~s$^{-1}$ at most (Lucy et al., in preparation). The 1990 optical high state thus remains a uniquely volatile incident in the recorded history of MWC 560. As we will show, there are still new insights to be gleaned from the 28-year-old observations of this event.

Setting the stage for our analysis, MWC 560's UV spectrum is usually obscured by a curtain of absorption by \ion{Fe}{ii} and similar ions \citep{Michalitsianos1991}. This iron curtain weakened dramatically by the end of 1990 April \citep{Skopal2005}, allowing us to see the outflow in a rare, unobscured state. When the next UV spectrum was obtained, on 1990 September 26, the \ion{Fe}{ii} absorption had returned with even more optical depth than before \citep{Maran1991}. The iron curtain was never again observed to vanish, even during {\it Swift} observations throughout the 2016 optical high state (Lucy et al., in preparation).

The paper is structured as follows: We describe the archival data (\autoref{data}). We show that a total disruption of the iron curtain in 1990 April unveiled high-ionization UV BALs, which were previously misidentified as \ion{Si}{ii} and \ion{Fe}{ii} (\autoref{results}). We discuss the physical consequences of this finding---including connections to quasars and novae---and the disappearance of the iron curtain, which has been optically thick at every other epoch (\hyperref[4.1]{\S~4.1}--\autoref{4.3}). The high-ionization BALs link MWC 560 with symbiotic star AS 304 ($\equiv$V4018~Sgr), and show that these objects are prototypes for an emerging class of BAL symbiotics (\autoref{4.4}). Last, we summarize our conclusions (\autoref{conclusions}).

All velocities and wavelengths for MWC 560 are heliocentric.


\section{Data} \label{data}

\afterpage{
\begin{table}
\caption{Selected IUE spectra of MWC 560 from \citet{Michalitsianos1991}, \citet{Maran1991}, and \citet{Bond1984}.}
\label{table1}
\begin{tabular}{llcc}
\hline
Date (UT) & IUE ID & Bandpass (\AA) & Resolution (\AA) \\
\hline
1984 Mar 10.95 & SWP22459 & 1150--1975 & 4--7 \\
1984 Mar 10.93 & LWP02920 & 1910--3300 & 5--8 \\
1990 Mar 14.98 & LWP17534 & 1910--3300 & 5--8 \\
1990 Mar 14.99 & SWP38361 & 1150--1975 & 4--7 \\
1990 Apr 29.90 & SWP38698 & 1150--1975 & 4--7 \\
1990 Apr 29.92 & LWP17832 & 1910--3300 & 5--8 \\
1990 Apr 29.94 & LWP17833 & 1845--2980 & 0.1--0.3 \\
1990 Apr 30.02 & SWP38699 & 1150--1975 & 4--7 \\
1991 Sep 29.12 & LWP21366 & 1910--3300 & 5--8 \\
1991 Sep 29.14 & SWP42580 & 1150--1975 & 4--7 \\
\hline
\end{tabular}
\end{table}
}

We retrieved NEWSIPS pipeline reductions of several {\it International Ultraviolet Explorer} (IUE) large-aperture spectra in the far-UV (FUV; 1150--1975\AA) and near-UV (NUV; 1910--3300\AA), including all UV spectra observed in 1990 April, from the Mikulski Archive for Space Telescopes (MAST). \autoref{table1} lists the spectra, which were previously published in \citet{Michalitsianos1991}, \citet{Maran1991}, and \citet{Bond1984}. 

\autoref{fig1} exemplifies the low-resolution (4--8\AA) spectra. The chosen dates follow Figure 1 in \citet{Skopal2005}, which was designed to exhibit the varying depths of MWC~560's iron curtain.

\section{Results: Line Identification} \label{results}

\begin{figure} 
\label{fig2}
\centerline{\includegraphics[width=3.45in]{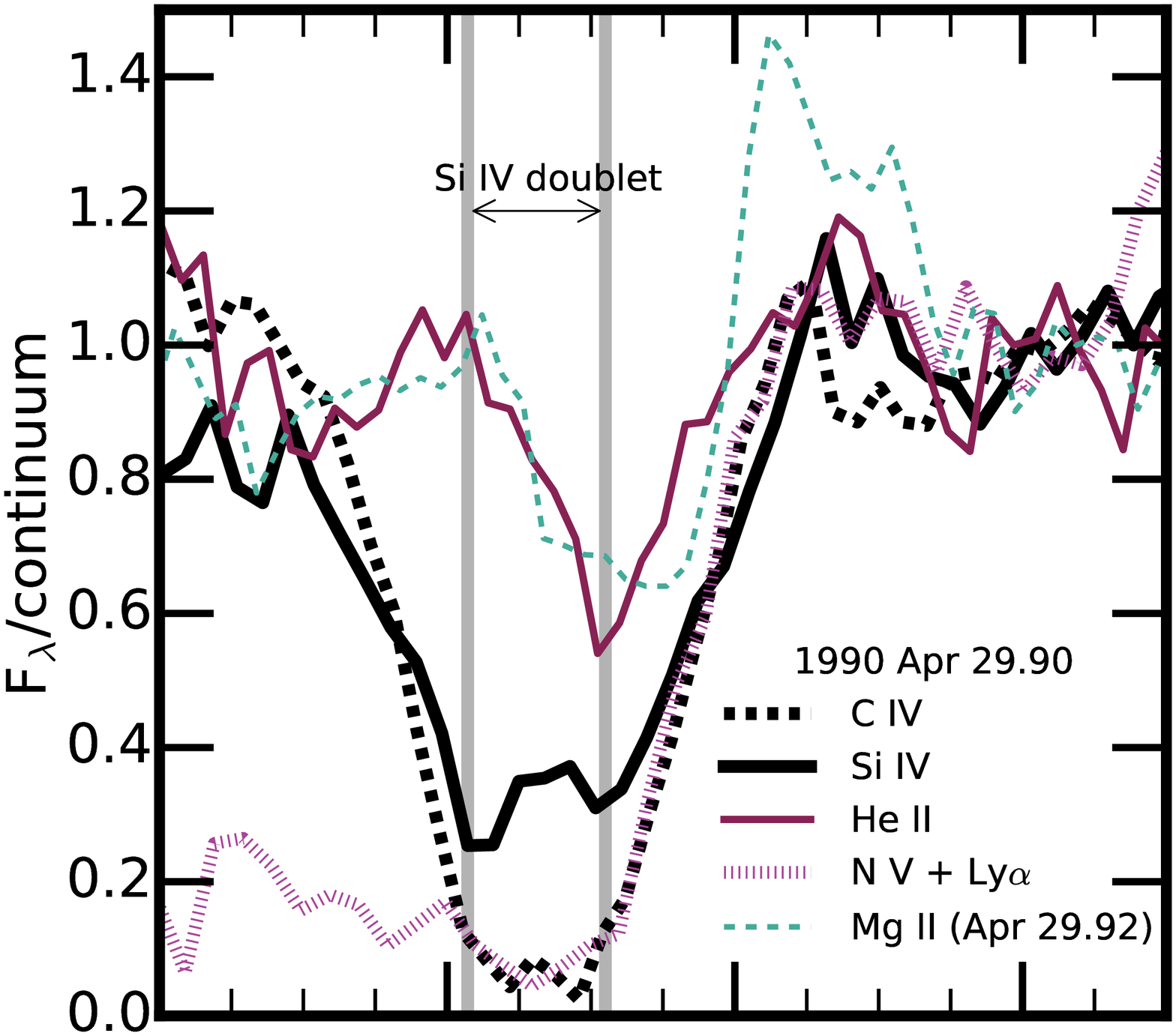}}
\centerline{\includegraphics[width=3.45in]{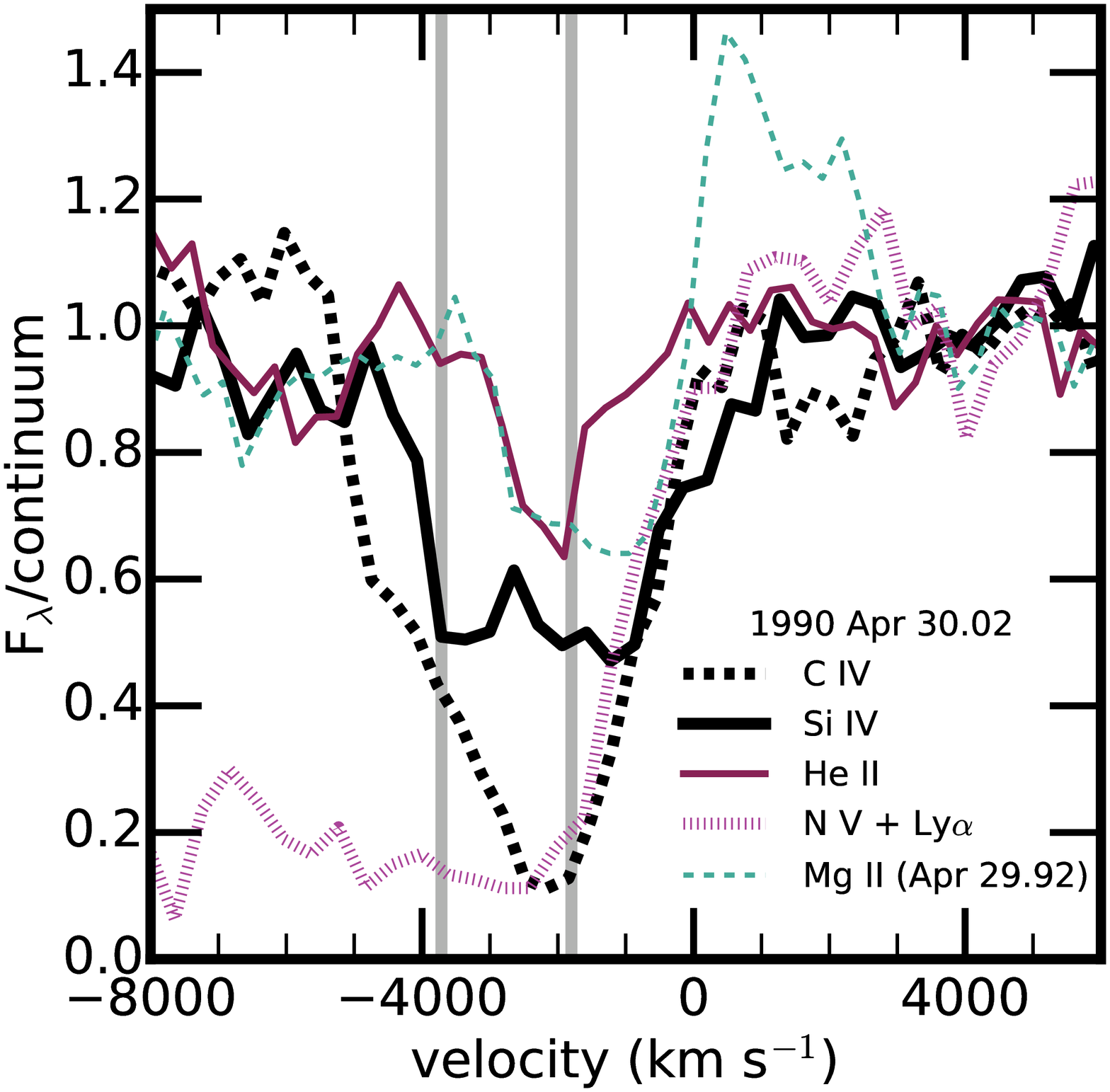}}
\caption{Velocity profiles from the high-ionization BALs in IUE spectra from 1990 April 29.90 (top) and 30.02 (bottom), and \ion{Mg}{ii} for comparison, with zero-velocity set to the rest wavelength of the reddest transition in each doublet or multiplet (\autoref{hibals}). The \ion{Si}{iv} doublet separation is marked at -1800~km~s$^{-1}$. Flux density is normalized to the median value between $+$3000 and $+$6000 km~s$^{-1}$. \vspace{0.4cm}} 
\end{figure}

\begin{figure*}
\includegraphics[width=\textwidth]{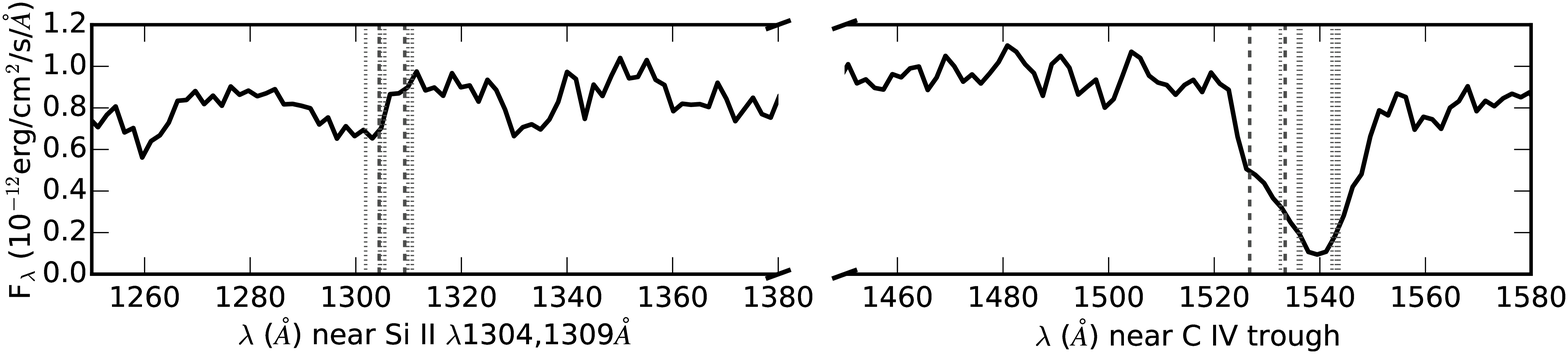}
\caption{The vertical lines drawn on this IUE spectrum from 1990 April 30.02 are \ion{Si}{ii} (dashed) and \ion{P}{ii} (hashed) transition rest wavelengths from \autoref{table2}. Atomic physics predicts that the two absorption complexes should be comparably strong, so the comparison region (left) shows that \ion{Si}{ii}$+$\ion{P}{ii} cannot account for the absorption trough near 1550\AA\ (right).} \label{fig3}
\end{figure*}

\subsection{High-ionization BALs} \label{hibals}

The deepest absorption features in the 1990 April~29.90 and 30.02 spectra are located at the expected positions of blue-shifted BALs from the high-ionization \ion{C}{iv}\,$\lambda$1548.20,\,1550.77\AA, \ion{Si}{iv}\,$\lambda$1393.76,\,1402.77\AA, and \ion{N}{v}\,$\lambda$1238.82,\,1242.80\AA\ doublets and \ion{He}{ii}\,$\lambda$1640\AA\ multiplet, as shown by rest-wavelength line-labels in \autoref{fig1}. The velocity profiles of these strong and rapidly variable BALs are shown in \autoref{fig2}. The \ion{Si}{iv} BAL is confirmed by the 9.0\AA\ (1900~km~s$^{-1}$) \ion{Si}{iv} doublet separation, which is about twice the local full-width-half-max spectral resolution and is observed in the trough (\autoref{fig2}). Subtracting the doublet separations, the \ion{C}{iv} and \ion{Si}{iv} troughs extend to at least 4000~km~s$^{-1}$. \ion{N}{v} is contaminated by Ly$\alpha$ absorption at high velocities. \ion{He}{ii} extends to about 3000~km~s$^{-1}$.

\subsection{Evidence against prior identifications} \label{3.2}

The feature we assign to \ion{C}{iv}\,$\lambda$1550\AA\ was previously misidentified as \ion{Si}{ii} near zero-velocity \citep{Michalitsianos1991,Maran1991} blended with \ion{P}{ii} \citep{Maran1991}. \autoref{table2} shows that there is a complex of \ion{Si}{ii} and \ion{P}{ii} transitions near 1304\AA\ with lower levels and oscillator strengths almost identical to those of \ion{Si}{ii} and \ion{P}{ii} near 1550\AA. \autoref{fig3} shows that the 1304\AA\ complex is only observed weakly and at large blue-shifts (if at all) in 1990 April. Therefore, \ion{Si}{ii} and \ion{P}{ii} cannot produce the deep trough near 1550\AA. Moreover, there are no \ion{Si}{ii} transitions from close to the ground state near the feature we assign to \ion{He}{ii}\,$\lambda$1640\AA, further contradicting line labels in \citet{Michalitsianos1991}. Our transition data are from the National Institute of Standards and Technology \citep[NIST;][]{NIST_ASD}; we checked for additional lines in \citet{kurucz}.

\begin{table}
\begin{centering}
\caption{NIST lab data showing that the 1304\AA\ and 1526\AA\ complexes of \ion{Si}{ii} and \ion{P}{ii} share similar line strengths: wavelength ($\lambda_{lab}$), lower-level excitation above the ground state (E$_{l}$), oscillator strength ({\it f}), and lower-level statistical weight (g$_{l}$).}
\label{table2}
\begin{tabular}{ccccc}
\hline
Ion & $\lambda_{lab}$ (\AA) & E$_{l}$ (eV) & {\it f} & g$_{l}$\\ 
\hline
\ion{Si}{ii} & 1526.71 & 0 & 0.133 & 2\\
\ion{Si}{ii} & 1304.43 & 0 & 0.093 & 2 \\
\\
\ion{Si}{ii} & 1533.43 & 0.036 & 0.133 & 4\\
\ion{Si}{ii} & 1309.27 & 0.036 & 0.080 & 4\\ 
\\
\ion{P}{ii} & 1532.53 & 0 & 0.008 & 1\\
\ion{P}{ii} & 1301.87 & 0 & 0.038 & 1\\
\\
\ion{P}{ii} & 1535.92 & 0.020 & 0.006 & 3\\
\ion{P}{ii} & 1536.42 & 0.020 & 0.002 & 3\\
\ion{P}{ii} & 1304.49 & 0.020 & 0.013 & 3\\
\ion{P}{ii} & 1304.68 & 0.020 & 0.009 & 3\\
\ion{P}{ii} & 1305.50 & 0.020 & 0.016 & 3\\
\\
\ion{P}{ii} & 1542.30 & 0.058 & 0.006 & 5\\
\ion{P}{ii} & 1543.13 & 0.058 & 0.001 & 5\\
\ion{P}{ii} & 1543.63 & 0.058 & 0.000 & 5\\
\ion{P}{ii} & 1309.87 & 0.058 & 0.010 & 5\\
\ion{P}{ii} & 1310.70 & 0.058 & 0.028 & 5\\
\hline
\end{tabular}\\
\end{centering}
\end{table}

The lines we assign to \ion{Si}{iv}\,$\lambda$1400\AA\ and \ion{N}{v}\,$\lambda$1240\AA\ were misidentified as \ion{Fe}{ii} \citep{Michalitsianos1991,Maran1991}. In modeling the 1990 March 14 iron curtain, \citet{shore1994} later speculated that \ion{Si}{iv} was probably present. In fact, the situation is clearer in the 1990 April spectra, in which most \ion{Fe}{ii} absorption has vanished. Any remaining weak lines would be from easily-populated \citep[see][]{lucy2014} excitations less than 0.4 eV above the ground state, there being no excitation states between 0.39~eV and 0.98 eV in the Fe$^{+}$ atom. We summed the relative optical depths of these low-excitation lines (oscillator strength $\times$ statistical weight $\times$ wavelength) into 10\AA\ bins to show that they cannot produce deep FUV absorption without producing deep NUV absorption, which is not observed in 1990 April (\autoref{fig1}). The only sign of iron-like absorption comes from a few of the strongest low-excitation lines (e.g., \ion{Fe}{ii}\,$\lambda$2600\AA) observed in the high-resolution NUV spectrum from 1990 April 29 as zero-velocity, saturated absorption $<$~1\AA\ wide; these narrow, possibly-interstellar lines do not produce deep features in the low-resolution spectra. Thus, although low-ionization lines may weakly contaminate the high-ionization BALs on 1990 April 29--30, they cannot account for them.

\section{Discussion} \label{disc}

\subsection{Ionization and velocity structure} \label{4.1}

Our results show that photo-ionization is likely a prominent mechanism in the outflow from MWC 560's accretion disk. The broad and continuous velocity distributions of BALs from a wide range of ionization states argue against collisional ionization in discrete shocks or a star-like photosphere alone. Ionizing photons of at least 33, 48, and 77 eV can populate the \ion{Si}{iv}, \ion{C}{iv}, and \ion{N}{v} BAL transitions, respectively; these are emitted in sufficient quantities at the $\sim$10$^{5}$ K inner radii of a WD accretion disk with MWC 560's parameters (\citealt{Schmid2001}).

The presence of high-ionization lines is consistent with the high-velocity absorption from \ion{He}{i}\,$\lambda$10830\AA\ in 1990 April 12 and 1991 January 25 infrared spectra of MWC 560 \citep{Meier1996}. This transition's metastable lower level is populated by recombinations of He$^{+}$; the ionization potentials of neutral helium and He$^{+}$ are 25 and 54 eV, respectively, so upward transitions from this state can be partially co-spatial with \ion{C}{iv} and \ion{Si}{iv} \citep[][]{leighly2011}.

The high-ionization \ion{C}{iv} and \ion{Si}{iv} BALs exhibit maximum radial velocities at least 1000~km~s$^{-1}$ faster than the contemporaneous low-ionization \ion{Mg}{ii}\,$\lambda$2800\AA\ resonance line (\autoref{fig2}). This behavior is also observed in BAL quasar outflows, for which various explanations have been proposed. Dense, self-shielding clumps may be embedded in a faster, higher-ionization gas \citep{Voit1993}. Equatorial winds can comprise vertically-stratified layers of ionization state and velocity \citep{Matthews2016}. Or an outflow may decelerate along the line of sight, with fast gas close to the photo-ionizing source shielding radially-slower gas closer to the observer \citep{Voit1993}. One or more of these geometries, which are not mutually-exclusive, could apply to MWC~560.

The \ion{He}{ii} BAL likewise appears to be slower than the \ion{C}{iv} and \ion{Si}{iv} BALs, probably for a different reason. This behavior is observed in CV and stellar winds, in which the \ion{He}{ii}\,$\lambda$1640\AA\ multiplet's highly-excited lower level (41 eV above the ground state) is thought to be collisionally populated in the hot, dense base of the outflow, where the gas is still accelerating \citep{hoare1994,smith2006}.

\subsection{Vanishing curtains and novae} \label{4.2}

Both ground-state and high-excitation Fe$^{+}$ vanished in 1990 April. The iron curtain has been present in every other UV spectrum of MWC 560, including observations in 1984 \citep{Bond1984}, 1990 January--March and late 1990--1993 \citep{Michalitsianos1991,Skopal2005}, 1995 (IUE PI: Starrfield), and throughout a 2016 high state that rivaled 1990 in bolometric luminosity (Lucy et al., in preparation), suggesting that the disappearance of Fe$^{+}$ signified more than a simple increase in ionizing luminosity.

Instead, the vanishing curtain probably corresponded to a temporary decrease in hydrogen column density---i.e., in total mass along the line of sight. In strongly photo-ionized plasma, iron is distributed between Fe$^{+2}$, Fe$^{+3}$, and Fe$^{+4}$, making it difficult to form much Fe$^{+}$ even through recombination \citep[see \S4.3 in][]{lucy2014}. Other `iron curtain' elements like silicon and chromium (but not magnesium\footnote{Although Mg$^{+}$ and Fe$^{+}$ have similarly small ionization potentials, Mg$^{+2}$ can dominate thanks to its large 80 eV ionization potential; recombination can then yield sufficient Mg$^{+}$ for opacity at 2800\AA\ \citep{lucy2014}, explaining the persistence of that BAL in MWC 560.}) have similar ionization potentials and behave in the same way. Iron curtain ions may survive only when enough mass is present to self-shield some portion of the outflow against high-energy photons.

The decline in column density during 1990 April was likely related to {\it (1)} the 1990 January--April optical brightening, which marked both a periodic peak in the system's light curve {\it and} a unique, non-periodic, and permanent tripling of the optical luminosity \citep{Leibowitz2015,Munari2016}; {\it (2)} the dramatic mass ejections of 1990 January--April, which yielded detached BALs blue-shifted up to a system record of 6000\,km\,s$^{-1}$ \citep{Tomov1990,Tomov1992}; and {\it (3)} the post-outburst, year-long suppresion of optical flickering and slowing of absorption speeds to less than 1000\,km\,s$^{-1}$ in late 1990--1991 \citep{Zamanov2011flickstop}. Coverage of the spectral response to these events is incomplete; to our knowledge, no optical spectra were obtained 1990 April~4 through October~22, and no UV spectra were obtained 1990 May~1 through September~25.

Although MWC 560 is unlikely to have ejected a nova-like shell\footnote{The optical photometry did not appear to fade substantially during 1990 April--May \citep{Tomov1990phot}, in contrast to the optical decline observed during the lifting of the iron curtain in novae \citep[e.g.,][]{Shoremanual}.}, it is nevertheless instructive to note that high-ionization UV BALs are briefly observed as P~Cygni profiles immediately after an iron curtain phase in some thermonuclear novae (\citealt{Shore1993}; M.~J.~Darnley 2018, private communication). The ejected mass rarefies as it expands, erasing the outer, heretofore-shielded, Fe$^{+}$ component. The similarity suggests that MWC 560's wind temporarily turned off in 1990 April after a series of powerful gusts, resulting in the nova-like phenomenology of a discrete mass ejection. This pattern supports the `discrete' and `quasi-stationary' taxonomy proposed by \citet{kolev1993} for MWC 560's optical absorption lines.

\subsection{A BAL nano-quasar} \label{4.3}

To our knowledge, there is no other accretion disk in the Galaxy observed to contain a similar range of ionization and excitation states in a persistent BAL outflow. There are CVs that exhibit both Balmer and \ion{C}{iv} BALs \citep{kafka2004}, but we have found no reports of any CV or X-ray binary known to have {\it also} exhibited \ion{Mg}{ii} and \ion{Fe}{ii} BALs except during thermonuclear novae. Unlike novae, Balmer BALs and the iron curtain are almost-permanent properties of MWC 560's spectrum for years at a time. Likewise, the high-ionization UV BALs may be present in every MAST-archived FUV spectrum of MWC 560, including the one obtained by \citet{Bond1984}, although contamination by the optically thick iron curtain precludes certainty at all epochs besides 1990 April.

Rather, MWC 560 best resembles Balmer FeLoBAL quasars, wherein photo-ionization models favor high densities in the BAL wind---comparable to densities in symbiotic nebulae (\autoref{intro}). These quasar winds persistently exhibit a similarly wide range of ionization states\footnote{Except for \ion{He}{ii}, whose absence in quasar outflows has been interpreted as a lower limit on the distance of absorbing material from the disk \citep{wampler1995}.} and similar spectra; see the quasar spectra in \autoref{fig1}. Moreover, iron curtain absorption has recently been observed to vary dramatically and even vanish in some FeLoBAL quasars, albeit on slower timescales than in MWC 560
(\citealt{rafiee2016}; see also \citealt{hall2011,mcgraw2015,zhang2015,stern2017}).

\subsection{The BAL symbiotics} \label{4.4}

The \ion{C}{iv}, \ion{Si}{iv}, and \ion{N}{v} BALs in MWC 560 illuminate a link to another symbiotic WD+RG binary: \citet{Munari1993} showed that AS 304 contains absorption blue-shifted up to 2200~$\pm$400~km~s$^{-1}$ from {\it only} \ion{C}{iv}, \ion{Si}{iv}, and \ion{N}{v}. Its FUV spectrum was observed only in 1992, but this observation was not prompted by any particular event, so there is no reason to think that the BALs were transient. The \ion{He}{ii} emission strength indicated a high luminosity on the order of 10$^{4}$ L$_{\odot}$ \citep{Munari1993}, which likely required nuclear burning on the surface of the white dwarf to power. The ionization parameter in AS 304's outflow was therefore large, contributing to the absence of low-ionization (and in this case, optical) absorption. 

In MWC 560, optical and UV flickering (\citealt{Bond1984,Tomov1996,Zamanov2011flick}; Lucy et al., in preparation; though see \citealt{Zamanov2011flickstop}) and hard X-rays sometimes observed from the boundary layer \citep{Stute2009} indicate an accretion disk without WD surface burning \citep[see][]{Luna2013}. Thus, both hot burning (AS 304) and cooler non-burning (MWC 560) symbiotics can sustain persistent BAL outflows.

MWC 560 and AS 304 have cousins in symbiotics with transient or low-velocity absorption. CH Cyg, a non-burning WD+RG binary with optical flickering, exhibited Balmer BALs with blue edges around 2000~km~s$^{-1}$ in 2017 January \citep{Iijima2017}; this absorption recurred sporadically throughout the year (\citealt{Teyssier2017}), sometimes appearing and vanishing within 1.5 hours (T. Iijima 2017, private communication). Z And, a burning-powered WD+RG binary, briefly exhibited a Balmer BAL blue-shifted up to $\approx$1700~km~s$^{-1}$ on 2006 July 9 \citep[Figure 7 in][]{Tomov2013}, near the peak of a 2 magnitude optical outburst. Meanwhile, many more symbiotics \citep{Tomov2013} and purported Be stars with cool giant companions \citep[including the iron stars; ][]{Cool2005} exhibit substantially narrower 100--1000~km~s$^{-1}$ absorption lines; we will not here attempt to disentangle their heterogeneous population, but some may mimic BAL outflow behaviors \citep{Tomov2013}.

The transient BAL outflows from symbiotics coexist with narrow jets, but the lines of sight to Z And and CH Cyg are severely misaligned from any plausible jet axis. Eclipses in CH Cyg indicate a large inclination angle \citep{Skopal1996} and the jet is extended in the plane of the sky \citep[][and references therein]{Weston2016}, suggesting that the jet cannot produce the BALs. We view Z And's orbit at an inclination angle of 47 $\pm$12$\degree$ \citep{Schmid1997}, and marginally resolved jets have been seen perpendicular to the orbital plane \citep{Broksopp} and 20$\degree$ west of the perpendicular's projection in the sky \citep{Sokoloski2006,kenny1995}. \citet{Burmeister2007} inferred a 1700~km~s$^{-1}$ jet during the 2006 outburst from narrow emission lines at $\pm$1150~km~s$^{-1}$, assuming that the jet was perpendicular to the orbit. Speculatively, the similarity between the jet and BAL velocity during this outburst might suggest that the opening angle of a jet-feeding wind briefly spread into the line of sight. 

Likewise, the degree to which MWC 560's outflow is collimated could vary with its strength. It probably is a generally polar outflow viewed at a small inclination angle\footnote{See discussion in \citet{Schmid2001} of unpublished echelle-spectrograph limits on TiO absorption band motion.}, but BAL quasars and CV winds show that it is problematic to constrain the degree to which a BAL outflow is collimated just from the ratio of absorption to emission equivalent widths \citep{Turnshek1997,Hamann1993,Voit1993,Drew1987}. BAL-producing outflows with wider opening angles than jets would help to explain the high incidence rate of BAL symbiotics, which now appear to be surprisingly well represented among the $\sim$10 \citep[e.g.,][]{Broksopp} symbiotics with known jets. Even with large half-opening angles up to $\pm$15$\degree$ \citep[as for the R Aqr jet;][]{Schmid2017}, less than 3.5\% of jets should point towards Earth---although a larger fraction might exhibit slower and shallower absorption projected into the line of sight.

\citet{Voit1993} proposed that low-ionization BAL quasars (cf. MWC 560, Z And in outburst, and CH Cyg) may represent a young evolutionary phase when accretion disk outflows entrain an ambient shroud of dust and gas, which disperses to lower column densities once exclusively high-ionization BALs are observed (cf. AS 304). The dense, frequently Fe$^{+}$-rich nebulae from cool-giant mass loss in symbiotic binaries \citep{shore1993b} resemble young quasar environments. We predict that higher-cadence FUV/optical spectral observations and larger sample sizes of both burning and non-burning symbiotics will lead to the detection of more BAL systems than narrow jets alone could produce, and open a window into the physics and geometry of quasar outflows on a nano-scale.


\section{Conclusions} \label{conclusions}

\begin{enumerate}[leftmargin=*,rightmargin=0ex,itemsep=0.8ex,label=\Alph{*}.]

\item  The FUV spectrum of symbiotic star MWC 560 contains high-ionization BALs from \ion{C}{iv}, \ion{Si}{iv}, \ion{N}{v}, and \ion{He}{ii} (\autoref{results}, \autoref{fig2}). They were unveiled most clearly on 1990 April 29--30, the only time throughout decades of UV observations that MWC 560's curtain of low-ionization iron absorption vanished (\autoref{fig1}).

\item The maximum radial velocities of the \ion{Mg}{ii} and \ion{He}{ii} BALs were at least 1000~km~s$^{-1}$ slower than contemporaneous \ion{C}{iv} and \ion{Si}{iv} BALs. These differences have precedents; \ion{Mg}{ii} is slower than \ion{C}{iv} and \ion{Si}{iv} in BAL quasars, and \ion{He}{ii} is slower than \ion{C}{iv} and \ion{Si}{iv} in CV and stellar winds (\autoref{4.1}).

\item The usually-persistent wind in MWC 560 temporarily switched to a discrete ejection phase in 1990. This finding is supported by the phenomenology of some novae, in which the iron curtain lifts to unveil high-ionization UV BALs (\autoref{4.2}).

\item Additional BAL symbiotics exist (\autoref{4.4}). The high-ionization BALs in MWC 560 illustrate a similarity to symbiotic star AS 304, which exhibits exclusively \ion{C}{iv}, \ion{Si}{iv}, and \ion{N}{v} BALs \citep{Munari1993}. Z~And and CH~Cyg have exhibited transient Balmer BALs \citep{Tomov2013,Iijima2017}; these transient BAL symbiotics contain jets, but the jets do not point towards Earth. Wide-angle disk winds are sometimes required to explain the BAL symbiotics, and may account for why the number of symbiotics with BALs is no longer negligible relative to the $\sim$10 symbiotics with jets.

\item MWC 560 and AS 304 most resemble Balmer FeLoBAL quasars and high-ionization BAL quasars, respectively (\autoref{4.3}, \autoref{4.4}). Both symbiotic star and quasar accretion disks are sometimes embedded in regions of dense gas and dust, so the evolution of these systems may involve similar physics.

\end{enumerate}


\section*{Acknowledgements}

We thank Matt Darnley for insights on connections to novae and editorial comments. ABL thanks Ulisse Munari, Karen Leighly, and Takashi Iijima for other edifying conversations. We thank Michael Rupen, Paul Kuin, Nirupam Roy, Gerardo Juan Manuel Luna, Jennifer Weston, and Peter Somogyi for editorial comments. ABL thanks the LSSTC Data Science Fellowship Program; their time as a Fellow has benefited this work. We are supported by NSF DGE-16-44869 (ABL), NSF AST-1616646 (JLS), and {\it Chandra} DD6-17080X (ABL, JLS). We gratefully acknowledge our use of pysynphot \citep{pysynphot}, PyAstronomy (\url{https://github.com/sczesla/PyAstronomy}), extinction \citep{extinction}, the Mikulski Archive for Space Telescopes (MAST), the International Ultraviolet Explorer (IUE), the Sloan Digital Sky Survey (SDSS), NASA's Astrophysics Data System (ADS), the SIMBAD database operated at CDS, and the National Institute of Standards and Technology (NIST).



\bibliographystyle{mnras}
\bibliography{lucy}




\bsp	
\label{lastpage}
\end{document}